\documentclass[11pt]{article}

\RequirePackage[a4paper,top=30.6mm,bottom=38.6mm,left=34.6mm,right=34.6mm,footskip=1.3cm]{geometry}
\usepackage{setspace}
\usepackage{amsmath}
\usepackage{amsfonts} 
\usepackage{amssymb}
\usepackage{graphicx}
\usepackage{hyperref}
\usepackage{tensor}
\usepackage{siunitx}
\usepackage{float}
\usepackage{bm}
\usepackage{slashed}
\usepackage{booktabs}
\usepackage{cancel}
\usepackage{multirow}
\usepackage{comment}
\usepackage{bbold}
\usepackage{caption}
\usepackage{amsmath}
\usepackage{enumerate}
\usepackage{cite}
\usepackage{tensor}
\usepackage{slashed}
\usepackage[utf8]{inputenc}
\usepackage{rotating}
\usepackage{bigfoot}

\usepackage[textsize=footnotesize]{todonotes}

\numberwithin{equation}{section}

\def \< {\left<}
\def \> {\right>}

\newcommand{\be}{\begin{equation}} \newcommand{\ee}{\end{equation}}
\newcommand{\bea}{\begin{eqnarray}}  \newcommand{\eea}{\end{eqnarray}}
\newcommand{\nn}{\nonumber}

\usepackage{amsfonts, amsthm}
\usepackage[english]{babel}
\usepackage{slashed}
\usepackage{mathrsfs}
\usepackage{amssymb}
\usepackage{color}

\begin{document}

	\begin{center}        
		\LARGE Emergence in String Theory and Fermi Gases
	\end{center}
	
	\vspace{0.7cm}
	\begin{center}        
		{\large  Jarod Hattab and Eran Palti}
	\end{center}
	
	\vspace{0.15cm}
	\begin{center}  
		\emph{Department of Physics, Ben-Gurion University of the Negev,}\\
		\emph{Be'er-Sheva 84105, Israel}\\[.3cm]
	\end{center}
	
	\vspace{1cm}
	
	
	\begin{abstract}
	\noindent  
	The Emergence Proposal suggests that some Swampland criteria, in particular on large field distances, are a consequence of the emergent nature of dynamics for fields in the infrared. In the context of type II string theory compactified on Calabi-Yau manifolds, it proposes that the cubic tree-level piece of the genus-zero prepotential is emergent from integrating out massive non-perturbative states. For a certain special non-compact Calabi-Yau, the blown-up conifold, it is known that the full all-genus prepotential can be matched onto the Grand Canonical potential of a two-dimensional Fermi gas. We propose here that this should be understood in the context of emergence: the prepotential is induced by integrating out the Fermi gas degrees of freedom. To make contact with the Swampland we need dynamical gravity, so compact Calabi-Yau manifolds. We show that for specifically the cubic term, an integrating out calculation also works for compact cases. In particular, the exact cubic term coefficient can be recovered  from integrating out a Fermi gas for any compact Calabi-Yau that is an elliptic fibration over a reflexive toric base. We also propose a general map, for any one-parameter Calabi-Yau, between the Grand Canonical potential of the ultraviolet non-perturbative system and the period. In particular, this map leads to an emergent cubic term in the genus-zero prepotential for any such one-parameter model. 
	\end{abstract}
	
	\thispagestyle{empty}
	\clearpage
	
	\tableofcontents
	
	\setcounter{page}{1}

\section{Introduction}
\label{sec:int}

It is expected that spacetime should be an emergent concept in quantum gravity, most probably from Planck sized constituents that are non-compressible, so satisfying a Pauli exclusion principle and Fermi-Dirac statistics \cite{tHooft:1993dmi}. Within an effective theory, the kinetic terms for fields involve spacetime derivatives and therefore, within this paradigm, should be emergent also. The Emergence Proposal \cite{Palti:2019pca}, based on the ideas in \cite{Harlow:2015lma,Heidenreich:2017sim,Grimm:2018ohb,Heidenreich:2018kpg,Palti:2019pca}, is an attempt to quantify this emergence of dynamics. It states that the dynamics of all fields arises due to integrating out towers of states down from a scale below the Planck scale. Simple toy-models of such integrating out suggest that emergence is manifest as constraints on the effective theory which match proposed Swampland constraints \cite{Harlow:2015lma,Heidenreich:2017sim,Grimm:2018ohb,Heidenreich:2018kpg,Palti:2019pca}. The proposal has been studied in some detail within string theory, see \cite{Grimm:2018cpv,Corvilain:2018lgw,Lee:2019xtm,Lee:2019wij,Palti:2020tsy} for a sample of some early work, and  \cite{Marchesano:2022avb,Castellano:2022bvr,vandeHeisteeg:2022btw,Cota:2022maf,Cribiori:2022nke,Marchesano:2022axe,Blumenhagen:2023tev,Castellano:2023qhp,Burgess:2023pnk,Baume:2023msm,Cribiori:2023ffn,Blumenhagen:2023yws,Blumenhagen:2023xmk,Seo:2023xsb,Calderon-Infante:2023ler,DeBiasio:2023hzo,Calderon-Infante:2023uhz, Castellano:2023aum,Castellano:2023jjt,Castellano:2023stg,Marchesano:2023thx,Basile:2023blg,Cota:2023uir,Hattab:2023moj,Casas:2024ttx,Blumenhagen:2024ydy}, for more recent investigations.

Within actual string theory models, emergence can be formulated as an exact infrared duality between the propagating fields and towers of non-perturbative states \cite{Grimm:2018ohb,Palti:2019pca}. This exact duality version, sometimes referred to as Strong Emergence, is difficult to check explicitly. Recently, some insights into this exact proposal were gained in the setting of type IIA string theory compactified on Calabi-Yau manifolds. Here the kinetic terms are controlled by an $N=2$ (genus-zero) prepotential ${\cal F}_0$, which is a function of the Kahler superfields $T^i$. It was known, since the work of Gopakumar and Vafa (GV) \cite{Gopakumar:1998ii,Gopakumar:1998jq}, that exponential (non-perturbative from the string worldsheet perspective) terms in the prepotential can be recovered by integrating out D2-D0 bound states using a Schwinger background field integral at strong string coupling. Strong emergence proposes that also the polynomial (tree-level) piece of the prepotential arises due to integrating out non-perturbative states \cite{Grimm:2018ohb}. 

The recent work in \cite{Hattab:2023moj} presented evidence for this by extending the Schwinger integral appropriately into the ultraviolet. This was done by analytically continuing the Schwinger proper time and including the pole at the origin (the ultraviolet) within the integration contour.\footnote{See also \cite{Blumenhagen:2023tev,Blumenhagen:2024ydy} for an alternative approach to addressing exact emergence.} This procedure maps to certain known integral representations of the periods for compact Calabi-Yau manifolds, and indeed reproduces the tree-level prepotential precisely. The conclusions of this approach were that the emergence of the tree-level piece of the prepotential comes from the deep ultraviolet, and that the degrees of freedom at that scale are point-like. Further, unlike the high multiplicity of the wrapped branes leading to the non-perturbative contributions, the tree-level piece comes from integrating out only a small number of independent degrees of freedom. So in the ultraviolet, the wrapped branes are resolved into a few constituents, which then need to be integrated out. 

In this work we present more evidence for this physics: that the tree-level prepotential arises from integrating out point-like constituents at the ultraviolet. We also clarify and sharpen some of the results in \cite{Hattab:2023moj}. The starting point is the work in \cite{Marino:2011eh}, which describes ABJM theory \cite{Aharony:2008ug} as a free Fermi gas in two dimensions. The Fermi gas determines the eigenvalues of a matrix model which arises from localisation of the ABJM theory on $S^3$. The same (matrix) model also describes topological strings on the blown-up conifold (a certain non-compact Calabi-Yau manifold) \cite{Aganagic:2002wv,Marino:2009jd,Drukker:2010nc}, which can be understood as a geometric transition, as in \cite{Gopakumar:1998ki}, from the ABJM theory on an $S^3$.\footnote{Note that in \cite{Marino:2011eh} the ABJM theory is studied holographically, so that the $S^3$ is on the boundary of AdS$_4$. While the geometric transition is such that the $S^3$ is in the internal Calabi-Yau.} In this work, in the context of emergence, we consider treating the Fermi gas not just as a model for the Matrix eigenvalues, but as a physical system of Fermions in two dimensions that is part of, or at least captures some of the physics of, the ultraviolet degrees of freedom. We will argue that this Fermi gas description should be thought of as capturing a certain aspect of emergence: that the prepotential arises from integrating out the degrees of freedom in the gas. This is discussed in section \ref{sec:intem}.

The free Fermi gas corresponds to topological strings on the blown-up conifold, that is a non-compact cone over a base $B=\mathbb{P}^1\times \mathbb{P}^1$. The Hamiltonian for the Fermi gas is mapped to the spectral curve of the topological strings. In \cite{Grassi:2014zfa}, this was generalised to other toric bases $B$. We review the key aspects of this work in section \ref{sec:buc}, with the particular example $B=\mathbb{P}^2$. In all known cases, this setup is always within a non-compact Calabi-Yau, where the four-dimensional gravity is non-dynamical (the Planck scale is infinite). We will propose how to extend these ideas to compact Calabi-Yau manifolds, with dynamical gravity, thereby allowing us to make connections with the Emergence Proposal and the Swampland. This is discussed in section \ref{sec:compa}.

\subsubsection*{Short summary of the results}

We can summarise our findings as follows. For the non-compact Calabi-Yau cones, there is a free Fermi gas description of ultraviolet degrees of freedom, and to it is associated a Grand Canonical potential $J\left(\mu\right)$, which is a function of the chemical potential $\mu$. We propose that in the context of emergence, the chemical potential is promoted to a dynamical degree of freedom, specifically mapped to a (vector-multiplet) modulus of type IIA string theory. The Grand Canonical potential is then understood as the effective action for $\mu$ coming from integrating out the Fermi gas degrees of freedom. In section \ref{sec:intfermigas}, we motivate this proposal with a simple toy model. 

There exists what we call a `quantum' map between the effective action for $\mu$ and the effective action for the type IIA modulus which appears in the prepotential. Such a map is known for the case of the blown-up conifold, and matches the Grand Canonical potential with the topological string Free Energy (all-genus prepotential) $F\left(T\right)$ exactly at the non-perturbative level \cite{Marino:2011eh,Hatsuda:2013oxa}
\be
T = \frac{4 \pi \mu}{\hbar} - i \pi \;,\;\;  J\left(\mu\right) = F\left(T\right)  \;.
\label{Jintomap}
\ee
Here $\hbar$ is the Planck constant of the Fermi gas description and is related to the inverse type IIA string coupling. The modulus $T$ is associated to the blow-up mode of the conifold.
Our proposed interpretation is therefore that the full non-perturbative all-genus prepotential is coming from integrating out the Fermi gas degrees of freedom.

For compact Calabi-Yau manifolds, there is no known map of the form (\ref{Jintomap}) from a Fermi gas Grand Canonical potential to an all-genus prepotential. However, we will take the approach that the physics we see for the blown-up conifold is general, being a manifestation of emergence, and so we propose that such a map does exist, at least for some compact cases, which captures the emergent nature of the effective action also when gravity is dynamical. More precisely, this map is from the Grand Canonical potential of some system of ultraviolet degrees of freedom, which does not need to behave necessarily as a free Fermi gas. 

We develop this map in two types of situations. First, when the compact Calabi-Yau is an elliptic fibration over a reflexive toric base $B$. Second, a general one-parameter Calabi-Yau.

For the first case, we have a modulus associated to the size of the fibre, which we denote $T_B$. Then we propose a Fermi gas model, and an identification which at leading order in $\mu$ takes the form
\be
T_B=\frac{2\pi\mu}{\hbar} + ... \;,\;\; J\left(\mu\right) = F\left(T_B\right)   \;.
\ee
The leading order approximation is sufficient to determine the (tree-level) cubic term in $T_B$ in the genus-zero prepotential, which is our primary interest in this paper. Indeed, it is because we are only after the cubic term that we can use this correspondence. The sub-leading terms are much harder to map in compact settings. From the Fermi gas perspective, we show that this approximation is such that the Fermi gas model captures only the high-energy modes of the ultraviolet system. The lower energy modes need not behave like a free Fermi gas. 

For the general one-parameter case, with modulus $T_B$, we propose the following. In the non-compact cone cases, it is known that there exists a `classical' map, to distinguish it from (\ref{Jintomap}), which maps the classical part of the Grand Canonical potential $J_0\left(\mu\right)$ to a period $\Pi_0(T_B)$ of the genus zero prepotential ${\cal F}_0\left(T_B\right)$ \cite{Grassi:2014zfa,Marino:2016new}:
\be
J_0(\mu) = \Pi_0\left(T_B\right) =  2{\cal F}_0\left(T_B\right)-T_B \frac{\partial {\cal F}_0\left(T_B\right)}{\partial T_B} \;.
\ee
We propose that such an identification is general, and holds also for compact Calabi-Yau manifolds. This can be done by definition: we can define the classical Grand  Canonical potential $J_0(\mu)$ for the ultraviolet system to be given by the period $\Pi_0(T_B)$. But for this to be meaningful we need to specify the map between $\mu$ and $T_B$. In section \ref{sec:spaintrel}, we propose such an exact classical map. The quantum map however is determined from the classical  map only at leading order in $\mu$, and takes the form 
\be
T_B = \frac{2 \pi r\mu}{\hbar} + ... \;,\;\; J\left(\mu\right) = F\left(T_B\right)   \;,
\ee
with $r$ some constant that can be determined for a given Calabi-Yau. The quantum map then gives us the interpretation, as in the blow-up conifold case, that the effective action for $T_B$ is emerging from integrating out the ultraviolet system, whether it behaves like a free Fermi gas or not.


This picture relates nicely to that of \cite{Hattab:2023moj} due to the classical part of the Grand Canonical potential $J_0(\mu)$ having a Mellin-Barnes contour integral representation, which is the same one as for the Calabi-Yau period $\Pi_0(T_B)$. This integral representation is the one which is interpreted as modified Schwinger integrals, and was argued to capture the integrating out procedure. In light of the work here, we can better understand it as capturing the integrating out to generate an effective action for $\mu$. Then there is a map from $\mu$ to the effective action for $T_B$. For the cubic term in the genus-zero prepotential, this does not really change anything. However, the exponential parts in the Mellin-Barnes integral are now understood to capture non-perturbative (in $g_s$) effects in the effective action, rather than the worldsheet instantons.

\section{Cones over toric bases and free Fermi gases}
\label{sec:buc}

In this section we summarise some of the relevant results in \cite{Marino:2011eh,Grassi:2014zfa}, to which we refer for the details. We consider type IIA string theory on certain non-compact Calabi-Yau threefolds. These are given as the total space of the fibration of the anti-canonical bundle $K_B$ over a two-fold base $B$. We can think of these as a cone over the base $B$. The base $B$ has some Kahler moduli parametrising its geometry. There are Kahler moduli parameterising its two-cycles. For example, for the case $B=\mathbb{P}^1\times\mathbb{P}^1$ (the blown-up conifold), there are two moduli parametrising the sizes of the two $\mathbb{P}^1$ factors. We will denote these type of moduli as $t^i$, and call them `local' Kahler moduli. They form part of superfields with complex scalar components $T^i = t^i + i b^i$, where the $b^i$ are periodic (pseudo-)scalars. Explicitly, we will consider only the case of a single such modulus $T$. So for the blown-up conifold we will set the size of the two $\mathbb{P}^1$'s equal.

There is also another special modulus which parameterises the size of the base $B$ as a whole. We will refer to this singled-out modulus as $t_B$, and the complexified superfield $T_B$, and call it a `global' Kahler modulus. A good way to think of it is through a (local) mirror configuration, of type IIB string theory, to the cone. This has only a single complex-structure modulus $U_B$, that is mirror to the Kahler modulus $T_B$.\footnote{We can think in terms of four-cycles Poincare dual to the two-form basis in which we expand the Kahler form. The base $B$ is the only such four-cycle in the cone. The two-cycles inside the base could potentially be completed to four-cycles in a compact setting, but need not be. That is why the mirror configuration has only a single modulus.} The Fermi gas model we consider will, in general, lead to an emergent cubic term in the prepotential for $T_B$.\footnote{Note that the cubic term is not well-defined geometrically for the non-compact case, in fact it can be defined through the Fermi model instead, but it is well-defined geometrically in the compact cases discussed in section \ref{sec:compa}.} However, we will initially primarily focus on the prepotential as a function of $T$. Of course, $T_B$ and $T$ are related, but it is important to clarify their distinction already at this stage. The distinction will be clearly marked with the subscript $B$ on the modulus.

The supergravity, and topological string, setting is well defined at weak string coupling $g_s \rightarrow 0$. The Fermi gas model is instead an expansion about strong string coupling $g_s \rightarrow \infty$. Here the semi-classical setting is controlled by a Planck's constant given by the inverse string coupling\footnote{Note that this is an analytic continuation in the string coupling.} 
\be
i\hbar = -\frac{\left(2\pi\right)^2}{g_s} \;.
\label{hbargs}
\ee
 The idea is that in this strong coupling description, the full (non-perturbative, all genus) prepotential is captured by the physics of a single two-dimensional free Fermion in a constant external potential. Concretely, the fermion is capturing the behaviour of eigenvalues of an exact Matrix model description of the system. In this description we have an expansion in $\frac{1}{g_s}$, while in the weakly-coupled description we have an expansion in $g_s$. 

As discussed, there are two relevant prepotentials: as a function of the `local' Kahler modulus $T$, and as a function of the `global' Kahler modulus $T_B$. We can extract both their forms from the Fermi gas description, but in different ways. In this section we will primarily discuss the `local' Kahler moduli, while in section \ref{sec:compa} we will discuss the `global' case.


\subsection{Statistical mechanics of Fermi gases}

In this section we review some basic notions of statistical mechanics of Fermi gases, so as to introduce the relevant quantities for the description. We are considering a two-dimensional free Fermi gas in Euclidean time.

The Grand Canonical potential can be written for a general system in terms of the one-particle Hamiltonian energy eigenstates as
\be
J\left(\mu\right) = \log \Big[ \sum_{\{n\}} e^{-\left(E_n - \mu N_n \right)} \Big]\;,
\ee
where $\{n\}$ is a given configuration $\{n_0,n_1,...\}$ where $n_i$ is the occupation number for the energy eigenstate with energy $\epsilon_i$. In our case, we would like to consider a Fermi gas in two dimensions, so $n_i=0$ or $1$. We defined the particle number $N_n=\sum_{i} n_i$ and total energy $E_n=\sum_{i} n_i \epsilon_i$. The parameter $\mu$ is the chemical potential. We can write this as
\be
J\left(\mu\right) = \mathrm{Tr} \log \left(1+e^{\mu-H}\right) =\int_{\epsilon_{\mathrm{min}}}^{\infty} \frac{n(E)}{e^{E-\mu}+1} dE \;.
\label{JintE}
\ee
Here $n\left(E\right)$ is the number of eigenstates with energy below $E$, so
\be
n\left(E\right) = \sum_{i}\Theta\left(E-\epsilon_i\right)\;,
\ee
with $\Theta$ the Heaviside step function. The minimum eigenvalue is $\epsilon_{\mathrm{min}}$, so that $n\left(\epsilon_{\mathrm{min}}\right)=0$.

There is a regime of the chemical potential where there exists a perturbative expansion of the Grand Canonical potential. That is the case when the minimal energy eigenstate is above the chemical potential
\be
 \epsilon_{\mathrm{min}}>\mu\;. 
 \label{orbreg}
 \ee
 In this case we can expand
\be
J(\mu) = - \sum_{l\geq1} \mathrm{Tr}\left[e^{-l H} \right] \frac{(-1)^l e^{l \mu}}{l}\;.
\label{Jorbifold}
\ee
In this expansion we can think of $l$ as counting the winding along the Euclidean time direction. The Fugacity is given by the expansion parameter $e^{\mu}$. 

\subsection{Hamiltonians for Calabi-Yau cones}
\label{sec:hamcycon}

In the Fermi gas picture, to calculate $J(\mu)$ we need to perform the integral (\ref{JintE}). We will primarily be interested in the classical approximation for $n(E)$. The Hamiltonian is a one-particle Hamiltonian in two-dimensions, so we have one momentum coordinate $p$ and one position coordinate $q$. So that once it is specified, the classical $n(E)$ is determined by the area
\be
n(E) = \int_{H(q,p)\leq E} \frac{dq dp}{2\pi\hbar} \;.
\label{neint}
\ee
Here $\hbar$ is the Planck constant of the Fermi gas theory, so that we are calculating an area in Planck units.

All that remains is to specify the Hamiltonian $H$. For a general toric base, this is given by \cite{Marino:2011eh,Grassi:2014zfa}
\be
e^{H} = \sum_{a=1}^{k+2} \mathrm{Exp}\left[ \nu_1^a q + \nu_2^a p + f^a \right]\;.
\label{speccur}
\ee
In the context of topological strings, this is known as the Spectral Curve.
Here the $f^a$ are constants, and will not be important for our purposes. The two constants $\nu_1^a$ and $\nu_2^a$ should be thought of as composing a set of two-dimensional vectors
\be
{\bf \nu}^a = \left( \nu_1^a, \nu_2^a\right) \;.
\label{vect}
\ee
The range of the index $a$ is determined by the integer $k$ (the number of linear relations of the vectors in the toric fan of the base). The vectors ${\bf \nu}^a$ are nothing but the vectors of the toric fan for the base $B$. For example, for the case $B=\mathbb{P}^2$, they are given by
\be
B=\mathbb{P}^2 \;:\; {\bf \nu}^1 = \left(\begin{array}{c} 1 \\ 0 \end{array}\right)\;, \;{\bf \nu}^2 = \left(\begin{array}{c} 0 \\ 1 \end{array}\right)\;, \;{\bf \nu}^3 = \left(\begin{array}{c} -1 \\ -1 \end{array}\right) \;.
\label{fanvp2}
\ee
This is illustrated in figure \ref{fig:p2tori}.

\subsection{The dictionary to topological strings}
\label{sec:topadic}

In this section, following \cite{Aganagic:2002wv,Marino:2009jd,Drukker:2010nc,Marino:2011eh,Grassi:2014zfa,Hatsuda:2013oxa}, we present the relation between the Fermi gas formalism and topological strings. We will consider the concrete case of the base being the blown-up conifold $B=\mathbb{P}^1\times\mathbb{P}^1$. This has two local Kahler moduli parameterising the two $\mathbb{P}^1$s, which we denote as $t_1$ and $t_2$. We consider the restriction $t_1=t_2=t$, and recall that $t$ is completed to a complex superfield $T=t+ib$. 

We consider the topological string Free Energy on this background $F(T)$. The claim is then that in the Fermi gas formalism this is mapped to the Grand Canonical potential \cite{Marino:2011eh}
\be
F\left(T\right) = J\left(\mu\right) \;,
\label{FtoJmap}
\ee
under the identification
\be
T = \frac{4 \pi \mu}{\hbar} -\pi i  \;.
\label{Tmumap}
\ee

It is informative to consider how the map (\ref{FtoJmap}) works \cite{Marino:2011eh}. We can expand the Grand Canonical potential in $\hbar$ as\footnote{Note that this expansion breaks the periodicity in $\mu \rightarrow \mu + 2 \pi i$. The claim is that the exact Grand Canonical potential is actually ${\cal J}(\mu)$ with $e^{{\cal J}\left(\mu\right)}=\sum_n e^{J\left(\mu + 2 \pi i n\right)}$. This is not important for our purposes.}
\begin{eqnarray}
J\left(\mu,\hbar\right) &=& \sum_{n=0}^{\infty} J_{n}(\mu) \left(2\pi\hbar\right)^{2n-1} + {\cal O}\left(e^{-\frac{4\pi\mu}{\hbar}} \right) \;.
\label{Jhbarexp}
\end{eqnarray}
The leading classical piece is $J_0(\mu)$. Note that the $J_n(\mu)$ themselves have polynomial parts and exponentials
\be
J_n(\mu) = J^{\mathrm{Poly}}_n(\mu) + {\cal O}\left(e^{-2\mu} \right) \;.
\ee
There is also a special role for the part in $J(\mu,\hbar)$ which is independent of $\mu$. The full exact expression can be written as
\begin{eqnarray}
    A(\hbar) = \frac{2\zeta(3)}{\pi\hbar}\left(1-\frac{\hbar^3}{16\pi^3}\right)+\frac{\hbar^2}{\pi^4}\int_0^{+\infty} dx\frac{x}{e^{\hbar x/\pi}-1}\log\left(1-e^{-2x}\right) \;.
    \label{indepmu}
\end{eqnarray}
For the Free Energy we have an expansion in string coupling
\be
F\left(T,g_s\right) = \sum_{g=0}^{\infty} {\cal F}_g(T) g_s^{2g-2} + {\cal O}\left( e^{-\frac{2 \pi i T}{g_s}}\right) \;.
\label{fhbarexp}
\ee
Similarly, we have polynomial and exponential terms at each genus
\be
{\cal F}_g(T) = {\cal F}_g^{\mathrm{Poly}}(T)+ {\cal O}\left( e^{-T}\right) \;.
\ee
The map (\ref{FtoJmap}) can be understood as holding under two identifications: the relation between $\hbar$ and $g_s$ (\ref{hbargs}), and the relation between $\mu$ and $T$ (\ref{Tmumap}).\footnote{Actually, the map (\ref{Tmumap}) is only exact perturbatively in $g_s$, in that it needs to be modified to capture the full non-perturbative contributions precisely. This is done through an effective Chemical potential $\mu \rightarrow \mu_{\mathrm{eff}}$ as in \cite{Hatsuda:2013oxa}.} 

Since the map between the Fermi gas and the topological string is a weak-strong duality, the perturbative expansions cannot be matched simply. However, there are two important things to note. First, the perturbative exponential terms are mapped to non-perturbative exponential terms. That is, worldsheet instantons are mapped to non-perturbative effects in $\hbar$, and perturbative exponentials in $\mu$ are mapped to non-perturbative terms in the string action. The latter map is quite subtle because the relation (\ref{hbargs}) involves an imaginary component, so that the direct map is $e^{-2\mu} \rightarrow e^{-\frac{2 \pi i T}{g_s}}$. This means we do not have a well-defined expansion for real small $g_s$ and large real part of $T$. Instead, the whole series of non-perturbative terms must be resummed into some function. By ${\cal O}\left( e^{-\frac{2 \pi i T}{g_s}} \right)$ in (\ref{fhbarexp}) we mean this function. The function can be calculated using the refined topological string \cite{Hatsuda:2013oxa}, and it behaves non-perturbatively in real $g_s$.\footnote{Note that the imaginary component in (\ref{hbargs}) does not lead to similar complications for the perturbative parts of the expansions.}

The second thing to note is that the genus-zero prepotential, which is the supergravity prepotential, has a special standing. This is because under supergravity it is protected by a non-renormalization theorem due to the dilaton being in the hypermultiplet sector. That means that we can see it in both the perturbative expansions. In the string coupling expansion it is simply defined as the part with $g_s^{-2}$ prefactor. In the Fermi gas $\hbar$ expansion it is the part multiplying $\hbar^2$. Note that the expansion (\ref{Jhbarexp}) is in odd powers of $\hbar$, but the map (\ref{Tmumap}) also involves a power of $\hbar$. So, for example, the cubic piece in $T$ in ${\cal F}_0$ comes from the cubic piece in $\mu$ in $J_0(\mu)$. While the linear piece in $T$ in ${\cal F}_0$ comes from the linear piece in $\mu$ in $J_1(\mu)$.


Let us make some comments to place the maps (\ref{FtoJmap}) and (\ref{Tmumap}) in a larger context. These are appropriate when thinking of the background $\mathbb{P}^1\times\mathbb{P}^1$ as a geometric spacetime, so in the large volume geometric regime. However, the Fermi gas describes all regimes. The orbifold regime, where the cycles are small, corresponds to $\mu$ being smaller than the minimal eigenvalue of the Hamiltonian. Another important point is that microscopically we can define these models through a geometric transition of Chern-Simons models with two integers: $N$ describing the rank of the gauge group and $k$ the Chern-Simons level \cite{Aganagic:2002wv,Marino:2009jd,Drukker:2010nc,Marino:2011eh}. For the $\mathbb{P}^1\times\mathbb{P}^1$ case the appropriate theory is ABJM \cite{Aharony:2008ug}. From this perspective, the spacetime background is quantized and the appropriate geometric parameter is the 't Hooft coupling $\lambda = \frac{N}{k}$. In the Fermi gas description $N$ is mapped to the particle number, and $k$ to Planck's constant $\hbar = \pi k$. The partition functions as functions of $T$ and of $\lambda$ are related by a Fourier transform \cite{Marino:2011eh}. In the Fermi gas, we work at small $\hbar$ and so in some sense give up the quantization of $k$. The spacetime quantization also disappears in the large $N$ limit, which is the geometric regime. From the Fermi gas perspective, this is the thermodynamic limit, where the partition function coincides with the Grand Canonical partition function. It is interesting to note that this spacetime quantization is studied from a different perspective as crystal melting in \cite{Okounkov:2003sp,Iqbal:2003ds,Ooguri:2009ijd,Ooguri:2009ri}.

\section{Emergence and integrating out}
\label{sec:intem}

In section \ref{sec:buc}, we reviewed a map between Fermi gas models and the partition function of topological strings. A part of this map included a map between the cubic term of the classical piece of the Grand Canonical potential and the cubic term of the genus zero prepotential. As discussed in the introduction, the Emergence Proposal claims that this tree-level term arises from an integrating out calculation. In this section we would like to approach the system from this perspective. 

Our proposal is that the map discussed in section \ref{sec:buc} is just an integrating out calculation: the effective action, as determined from the topological string partition function, for the modulus $T$ is arising from integrating out the Fermi gas degrees of freedom. 

We will argue for this below by performing an integrating out calculation of a two-dimensional field theory with a free Fermion placed at each point in a four-dimensional spacetime. Crucially, this will be a field theory, so second-quantized, integrating out procedure. We should really think of this as a toy-model for a full M-theoretic second-quantized treatment of the ultraviolet degrees of freedom. In such a full treatment, the full spacetime would also be emergent.\footnote{Note that, as discussed in section \ref{sec:hamcycon}, in the Chern-Simons geometric dual picture, the particle number $N$ is mapped to the number of branes. Therefore, a second-quantized treatment of the Fermi gas is analogous to a string field theory calculation.} 

\subsection{Integrating out the Fermi Gas}
\label{sec:intfermigas}

As discussed, we are taking a simplified approach to the integrating out calculation where we perform a naive integrating out of the Fermions directly in a field theory. The theory that we are proposing to integrate out is of a free Fermion in two dimensions with non-canonical kinetic terms and a spatially-dependent potential. However, to see the relation between the map and integrating out let us consider a simpler theory of a free Fermion with no potential and canonical kinetic terms. 

We consider a Fermion in six (Euclidean) dimensions with Euclidean action
\be
S^{(6)}_{\psi} = \int  d^4x d^2y {\cal L}^{(2)}_{\psi} \;.
\label{6daction}
\ee
The two dimensions are topologically a circle, with coordinate $\tau$, times a line, with coordinate $q$, so $d^2y = d\tau dq$.
The two dimensional Euclidean Lagrangian density is\footnote{The Euclidean derivative is defined such that $i\slashed{\partial}_E = \gamma^0 \partial_{\tau}-i\gamma^1 \partial_{q}$.}
\be
{\cal L}^{(2)}_{\psi} = \bar{\psi}_x(y) i\slashed{D}_E \psi_x(y) + m \;\bar{\psi}_x(y)\psi_x(y)\;.
\label{2dlag}
\ee
Here $m$ is the mass of the Fermion.
The two-dimensional Lagrangian involves two-dimensional fermions and only kinetic terms along the two dimensions. The subscript $x$ on $\psi_x$ is to remind us that there is a field at every point in the $x$ directions, but since there are no kinetic terms along those directions we just have a copy of the system at each point in $x$.

  The gauge covariant derivative couples the fermions to a non-propagating background gauge field, and we give it a background along the $\tau$ direction which is constant along the $y$ directions but can vary over the $x$ directions: $A_{\tau}=\mu(x)$, so that we can write the Lagrangian as
\be
{\cal L}^{(2)}_{\psi} = \bar{\psi}_x(y) i\slashed{\partial}_E \psi_x(y) - \mu\left(x\right) j_0 + m \;\bar{\psi}_x(y)\psi_x(y) \;,
\ee 
with $j_0 = \bar{\psi}_x\gamma^0\psi_x$. So this system describes a two-dimensional theory of a Fermion, coupled to a background gauge field, at each point in the four-dimensional spacetime.

Now we wish to integrate out the two dimensional system, to arrive at an effective four-dimensional action for $\mu(x)$. We can write the partition function
\be
Z = \int \mathcal{D}i\psi^{\dagger}{\cal D}{\psi} e^{-S^{(6)}_{\psi}} \;.
\ee
Performing the integrating out calculation is standard, see for example \cite{Kapusta_Gale_2006}. Let us define the partition function at each point $x$ as $Z_x$ such that $Z = \prod_x Z_x$.
Then we have for this partition function
\be
Z_x=\int_{\text{anti-period}}\mathcal{D}i\psi_x^{\dagger}\mathcal{D}\psi_x \exp \left[\int_0^{1} d \tau \int dq \; \bar{\psi}_x\left(-\gamma^0 \frac{\partial}{\partial \tau}+i\gamma^1 \partial_q-m+\mu \gamma^0\right) \psi_x\right] \;.
\ee
For each point in $x$ we can expand the fermion as
\begin{eqnarray}
    \psi_{\alpha}(\tau,q) = \frac{1}{\sqrt{V}}\sum_n\sum_{p}e^{i\left(pq+\omega_n\tau\right)}\Tilde{\psi}_{\alpha,p,n} \;,
\end{eqnarray}
with $\omega_n = 2\pi(n+\frac{1}{2})$. Here $\alpha$ is the spinor index, and $V$ is the one-dimensional space regularising volume. We have also discretised the momentum $p$ along the $q$ direction.
We can then write the partition function as
\be
Z_x=\left[\prod_n \prod_{p} \prod_\alpha \int id \tilde{\psi}_{\alpha,p, n}^{\dagger} d \tilde{\psi}_{\alpha,p, n}  \right] e^{-S} \;,
\ee
where
\be
-S =\sum_n \sum_{p} i \tilde{\psi}_{\alpha,p, n}^{\dagger} D^{\alpha\rho} \tilde{\psi}_{\rho,p, n} \;. 
\ee
The operator $D$ is a matrix
\be
D =-i \left[\left(-i \omega_n+\mu\right)-\gamma^0 \gamma^1 p-m \gamma^0\right] \;.
\ee
We therefore have
\be
Z_x=\operatorname{det} D \;.
\ee
The determinant is carried out over both Dirac indices and in frequency-momentum space. Using
\be
\ln \operatorname{det} D=\operatorname{Tr} \ln D \;,
\ee
one finds that
\be
\ln Z_x= \sum_n \sum_{p} \log \left[\left(\omega_n+i \mu\right)^2+\omega_p^2\right] \;,
\ee
with $\omega_p = \sqrt{p^2+m^2}$. Because the sum is over $n \in \mathbb{Z}$ we can replace
\be
\log \left[\left(\omega_n+i \mu\right)^2+\omega_p^2\right] \rightarrow \frac12\ln \left[(\omega_n^2+\omega_p^2-\mu^2)^2+4\mu^2\omega_n^2\right] \;,
\ee
and then factorising we have
\begin{eqnarray}
\ln Z_x = \frac12\sum_n \sum_{p} \left[\ln \left(\omega_n^2+(\omega_p-\mu)^2\right)+\ln \left(\omega_n^2+(\omega_p+\mu)^2\right) \right]\;.
\end{eqnarray}
To evaluate the sum, we first write
\be
\ln \left[\omega_n^2 +(\omega_p \pm \mu)^2\right] =  \int_1^{(\omega_p \pm \mu)^2} \left(\frac{1}{\omega_n^2+\theta^2} \right)d \left(\theta^2\right) +\ln \left(1+\omega_n^2\right) \;.
\ee
The sum over $n$ can be carried out by using the summation formula
\be
\sum_{n=-\infty}^{\infty} \frac{1}{\omega_n^2+\theta^2}=\frac{1}{\theta}\left(\frac{1}{2}-\frac{1}{\mathrm{e}^\theta+1}\right) \;.
\ee
Integrating over $\theta$, we finally obtain
\be
\ln Z_x= V \int \frac{dp}{2 \pi}\left[\omega_p+\ln \left(1+\mathrm{e}^{-\omega_p+\mu}\right)+\ln \left(1+\mathrm{e}^{-\omega_p-\mu}\right)\right] \;.
\ee
We can write
\be
V \int  \frac{dp}{2 \pi} = \int\frac{dq dp}{2 \pi} \rightarrow \mathrm{Tr} \;,
\ee
and so have for the full partition function
\be
\log Z  = \int d^4x \;\mathrm{Tr}\Big[ H_{\psi} + \log\left(1+e^{-\left(H_{\psi}-\mu \right)} \right) + \log\left(1+e^{-\left(H_{\psi}+\mu \right)} \right)\Big]  \;.
\ee
Note that here the trace does not include the Kaluza-Klein modes along $\tau$, they have already been summed over. The Hamiltonian $H_{\psi}$ is the one-particle Hamiltonian for the Fermion $\psi$. 

We can therefore write an effective four-dimensional action for $\mu(x)$ as
\begin{eqnarray}
S^{(4)}_{\mathrm{eff}}(\mu) &=& -\int d^4x \; \mathrm{Tr}\Big[ \log\left(1+e^{-\left(H_{\psi}-\mu \right)} \right)\Big] \nonumber \\
&-& \int d^4x \;\mathrm{Tr}\Big[ H_{\psi}  + \log\left(1+e^{-\left(H_{\psi}+\mu \right)} \right)\Big] \;.
\label{effmuac}
\end{eqnarray}
This should be compared with the expression for the Grand Canonical potential (\ref{JintE}). We see that the first term in (\ref{effmuac}) then gives us the relation
\be
S^{(4)}_{\mathrm{eff}}(\mu) =-\int d^4x \;J(\mu(x)) \;.
\label{seffJmu}
\ee
The other contributions in (\ref{effmuac}) will not be present in a full supersymmetric calculation. This is because there is no vacuum energy, and we only count the particles and not the anti-particles in the BPS states.

The calculation presented justifies an understanding of the map between the Grand Canonical potential and the effective action for $\mu$. In the actual Fermi gas model there are two main differences relative to the calculation presented. The first is that there is spatially dependent potential for the Fermion. The second is that the kinetic terms are not canonical. While these may give rise to technical complications, they do not change anything conceptually. In fact, the potential implies that the trace in (\ref{effmuac}) becomes better defined. We can also map a free Hamiltonian with an arbitrary kinetic term to a free field with the same kinetic term under the usual map of the momentum to the spatial derivative. 

In summary, we have presented a toy model simple calculation which motivates our proposal: that the Grand Canonical potential should be understood as the effective action for the chemical potential upon integrating out the Fermi gas.

\subsection{Topological string partition function and supergravity}
\label{sec:backgrsusy}

In section \ref{sec:intfermigas} we presented a calculation which showed that we should think of the Grand Canonical potential as the effective Lagrangian density for the chemical potential $\mu$ after integrating out the Fermi gas. Our proposal is that this is precisely capturing the map (\ref{FtoJmap}), and so we should think of that map as integrating out. The idea is that $F(T)$ is capturing the effective action in a self-dual graviphoton background. More precisely the effective action is most efficiently written in superspace, with Grassmann components $\theta$. The relevant term reads (see for example \cite{Dedushenko:2014nya})
\be
S^{(4)}_{\mathrm{eff}} = -i\int d^4xd^4\theta F(X^{I},W) \;.
\label{sugF}
\ee 
Here the $X^I$ are vector superfields, with scalar components $T^I$ 
\be
X^I = T^I + {\cal O}(\theta) \;,
\ee
and $W$ is the graviphoton superfield.\footnote{This should be understood as derived in a background for the whole graviphoton superfield.} So we are proposing that the result (\ref{seffJmu}) should be equated with (\ref{sugF}), in a complete treatment rather than our simplified analysis. Of course, this identification involves the quantum map from $\mu$ to $T$ (and $\hbar$ to $g_s$).\footnote{It would be interesting to understand if there is a meaning to the effective action at strong $g_s$.}

We can expand
\be
F(X^{I},W) = \sum_{g=0}^{\infty} {\cal F}_g(X) W^{2g} \;.
\ee
This expansion should be identified with the topological string genus expansion of the Free energy
\be
F(T) = \sum_{g=0}^{\infty} {\cal F}_g(T) g_s^{2g-2} \;.
\ee
In this sense, we have recovered the identification $F(T)=J(\mu)$, as in (\ref{FtoJmap}), from the integrating out analysis leading to (\ref{seffJmu}). 

Note that we are particularly interested in the term in (\ref{sugF}) which gives rise to the tree-level kinetic terms for the graviphoton and the vector multiplets. This is contained in the $g=0$ piece \cite{Dedushenko:2014nya}
\be
S^{(4)}_{\mathrm{eff}} \supset -i\int d^4xd^4\theta {\cal F}_0(X) \;. 
\ee
More precisely, at large volumes $T \rightarrow \infty$, the kinetic terms are coming from the cubic terms in ${\cal F}_0(T)$.

Actually, to be precise, the analysis in section \ref{sec:topadic} is for a non-compact setting, in which case the supergravity approach is not well-defined. However, we now proceed to argue that similar physics holds also in compact Calabi-Yau settings.

\section{Emergence at infinite distance in compact models}
\label{sec:compa}

In section \ref{sec:topadic}, we saw that for the blown-up conifold $B=\mathbb{P}^1\times\mathbb{P}^1$ we have an exact non-perturbative map, given by (\ref{FtoJmap}) and (\ref{Tmumap}), between the all-genus topological string prepotential $F(T)$ and the Fermi gas Grand Canonical potential $J(\mu)$. In section \ref{sec:intem}, we argued that this map should be understood as an integrating out calculation. The aim of this section is to generalise some of these results also to compact Calabi-Yau manifolds, where the four-dimensional gravity is dynamical. 

For the blown-up conifold, the Fermi gas model was extremely powerful, yielding even non-perturbative terms. We have no realistic ambitions to reproduce this type of accuracy for compact models. Instead, we focus only on the tree-level cubic term in the genus-zero prepotential. This is the most interesting term from the perspective of Emergence. The crucial point is that this term is also leading and classical from the Fermi gas side, so at strong coupling, being associated to the cubic term in $\mu$ in the classical Grand Canonical potential $J_0(\mu)$. This means that in proposing a Fermi gas-type picture for compact models, we only need to specify the physics which, upon integrating out, leads to the leading classical part. This physics may have a much simpler description than the full ultraviolet physics.

We apply this possible simplification in the context of compact elliptic fibrations of reflexive toric bases $B$ in section \ref{sec:cubco}. In section \ref{sec:spaintrel} we consider instead a general one-parameter compact Calabi-Yau. In both these cases, we will present the leading order calculation which yields the cubic part of the genus-zero prepotential as emergent in the same sense as discussed in sections \ref{sec:buc} and \ref{sec:intem}.

\subsection{The classical map for the global modulus}
\label{sec:glob}

For the blown-up conifold, the map (\ref{FtoJmap}) and (\ref{Tmumap}) involves the modulus $T$, which is a `local' modulus, in the language of section \ref{sec:buc}. In order to generalise the setting to compact models, we first need to understand how to calculate such a map for the `global' modulus $T_B$. This is essentially the mirror modulus to the complex-structure modulus associated with the spectral curve. It captures the size of a curve, but one that is in some sense `universal' to the base. In fact, no such exact quantum map is known for this base-associated modulus $T_B$. However, there is a weaker result, that we term a `classical' map, which we can utilise to obtain an approximate quantum map that will be sufficient for our purposes. 

For the non-compact cones, there is an exact map between the leading classical piece of the Grand Canonical potential $J_0(\mu)$ and the genus-zero prepotential that goes through the associated period $\Pi_0(T_B)$ as\cite{Grassi:2014zfa,Marino:2016new}
\be
J_0(\mu) = \Pi_0(T_B) = 2{\cal F}_0\left(T_B\right)-T_B \frac{\partial {\cal F}_0\left(T_B\right)}{\partial T_B} \;.
\label{j0f0rel}
\ee 
This holds under an analytic continuation of $\mu$, as in (\ref{Tmumap}), and an identification 
\be
U_B = r \mu \;,
\label{tbride}
\ee
with $U_B$ being the mirror (complex-structure) modulus to $T_B$. 
The integer $r$ is associated to the embedding of $B$ in the Calabi-Yau cone and is given for different cases in \cite{Grassi:2014zfa}. The relation between $U_B$ and $T_B$ is in general complicated, given by the mirror map, but at leading order in large volume is
\be
T_B = U_B + {\cal O}\left(e^{-U_B}\right) \;.
\label{tbubmap}
\ee
We call the relations (\ref{j0f0rel}) and (\ref{tbubmap}) (or more precisely (\ref{tbride})), the classical map.

We are primarily concerned with the leading cubic term in the genus-zero prepotential. We note that for that term, the map (\ref{j0f0rel}) and (\ref{tbubmap}) implies the relation
\be
T_B = r \mu \;,\;\; \left. J_0(\mu) \right|_{\mathrm{cubic}} = -\left. {\cal F}_0\left(T_B\right) \right|_{\mathrm{cubic}} \;.
\label{j0f0cubic}
\ee
This looks like a piece of a more general relation similar to the one for the blown-up conifold (\ref{FtoJmap}). We therefore propose that such a map also exists for the global modulus $T_B$, 
\be
J\left(\mu\right) = F\left(T_B\right) \;,
\label{TtoFglobal}
\ee
with an identification that at large base volume, $T_B\rightarrow \infty$, takes the leading form 
\be
T_B = \left(\frac{2 \pi }{\hbar}\right) r \mu + ... \;. 
\label{ttomuglobal}
\ee

The proposed leading order behaviour of the quantum map (\ref{ttomuglobal}) reproduces the correct cubic term as in (\ref{j0f0cubic}). It also matches at leading order the exact map in the local modulus $T$ for the blown-up conifold (\ref{Tmumap}) (which has $r=2$), under the identification $T_B=T$, which is correct given that the volumes of the two $\mathbb{P}^1$'s are set equal and so determine directly the volume of the full base $B$. 

Further strong evidence for the proposed procedure in going from the classical map (\ref{j0f0cubic}), to the leading-order quantum map (\ref{ttomuglobal}), comes from considering the linear terms in $T_B$ in the genus-zero ${\cal F}_0$ and genus-one ${\cal F}_1$ prepotentials. Let us define the linear terms
\bea
\left.J_0(\mu)\right|_{\mathrm{linear}} &=& l_{J_0} \;\mu \;\;, \nonumber \\
\left.{\cal F}_0(T_B)\right|_{\mathrm{linear}} &=& l_{{\cal F}_0} \;T_B \;\;, \nonumber \\
\left.{\cal F}_1(T_B)\right|_{\mathrm{linear}} &=& l_{{\cal F}_1} \;T_B \;\;,
\eea
with $l_{J_0}$, $l_{{\cal F}_0}$ and $l_{{\cal F}_1}$ being some constants. Then the classical map (\ref{j0f0rel}) gives the relation
\be
l_{J_0} = r\; l_{{\cal F}_0} \;,
\label{classrel}
\ee
which fixes $l_{J_0}$. Once this is fixed, we can determine the quantum map as relating to the linear term in the genus-one prepotential. The quantum map (\ref{ttomuglobal}) yields the relation
\be
l_{J_0} = 4 \pi^2 r \;l_{{\cal F}_1} \;.
\label{quantrel}
\ee 
Comparing (\ref{classrel}) and (\ref{quantrel}) we see that the quantum map predicts the relation
\be
l_{{\cal F}_0} = 4 \pi^2 \;l_{{\cal F}_1}\;.
\ee
Indeed, this relation holds for Calabi-Yau manifolds, where the linear terms takes the form\footnote{Note that this form is symplectic basis dependent, and holds for the canonical large volume or large complex-structure basis.}
\be
l_{{\cal F}_0} = \pi^2\frac{c_2 \cdot \omega_B}{6} \;,\;\; l_{{\cal F}_1} = \frac{c_2 \cdot \omega_B}{24}\;,
\ee
with $c_2$ being the second Chern class of the Calabi-Yau and $\omega_B$ being the two-form associated to the Kahler modulus $T_B$.

In practice, we can determine the cubic term in the genus-zero prepotential from the Grand Canonical potential directly from the classical map (\ref{j0f0cubic}). The reason for the proposal that this is part of a more general relation (\ref{TtoFglobal}), with a leading-order quantum map (\ref{ttomuglobal}), is required to conceptually connect it to integrating out, since integrating out leads to the prepotential and not the period. Indeed, a similar issue in difference between the prepotential and the period arose in the spacetime integrating out picture of \cite{Hattab:2023moj}, which we discuss in section \ref{sec:spaintrel}.

\subsection{The cubic term at large $\mu$ for the non-compact cones}
\label{sec:cubnonco}

The relation (\ref{j0f0cubic}) determines the cubic term in the genus-zero, so supergravity, prepotential from the Fermi gas model. We have argued that it should be understood as a consequence of emergence. It means that we should be able to calculate the coefficient of the cubic piece from  the Fermi gas picture. In this section we outline this calculation for the non-compact cones case, which is a review of results in \cite{Marino:2011eh,Grassi:2014zfa}. In section \ref{sec:cubco}, we will argue that a similar calculation holds also for compact cases.

Our starting point is considering a cubic term in the prepotential for $T_B$
\be
{\cal F}_0\left(T_B\right) = -\frac{\kappa_B}{6}\; T_B^3 + ... \;,
\ee
where $\kappa_B$ is what we would call the triple intersection number in the compact case (in the non-compact case this is not well-defined geometrically). Using the relation (\ref{j0f0cubic}), we therefore have 
\be
J_0(\mu) = \frac{r^3\kappa_B}{6}  \mu^3 + ... \;.
\ee
In turn, this can be traced back to the behaviour 
\be
2 \pi \hbar\; n(E) = \frac{r^3\kappa_B}{2} E^2 + ... \;.
\label{tripnE}
\ee

\begin{figure}
\centering
 \includegraphics[width=0.45\textwidth]{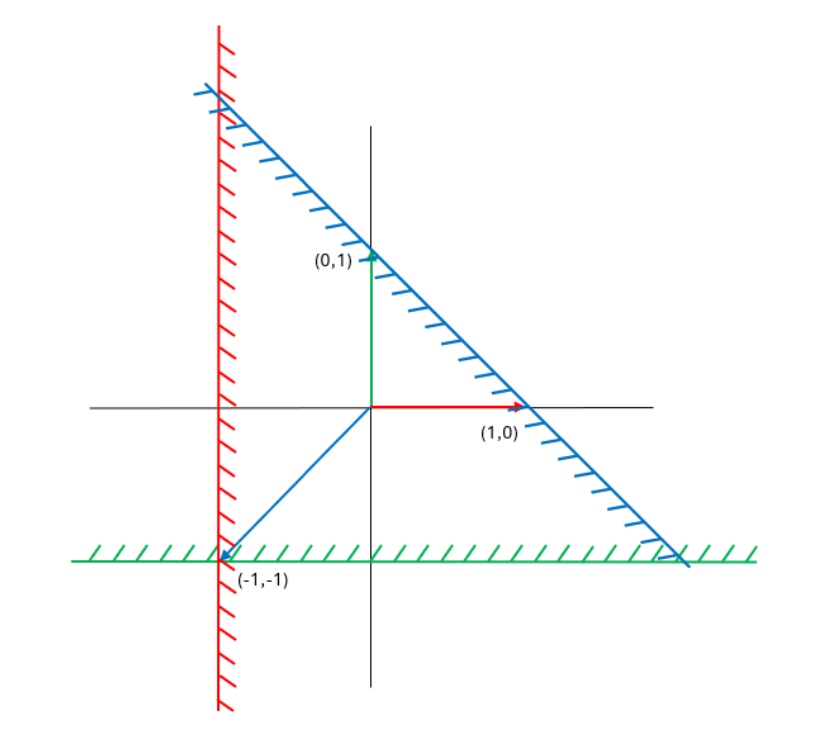}
\caption{Figure showing the toric fan and dual polygon for $B=\mathbb{P}^2$. The vectors of the fan are given in (\ref{fanvp2}). For each vector, the dual area is the set of points satisfying $\left<u,{\bf \nu}^a \right> \geq -1$, which are illustrated by the same colour line and the hatching denoting the direction of the points satisfying the inequality. The overlap of the area for all the vectors is the area of the (dual) polygon. In this case, being $\frac{9}{2}$.}
\label{fig:p2tori}
\end{figure}

We have that $n(E)$ is given by the area formula (\ref{neint}). We are interested in calculating this area in the high energy limit $E \rightarrow \infty$. This is the so-called tropical limit of the curve (\ref{speccur}), in which case (\ref{neint}) is given by the area of a polygon delimited by the lines\footnote{More precisely, the tropical limit is $\nu_1^a q + \nu_2^a p + f^a = E$, but the $f^a$ parameters are not important in the limit $E \rightarrow \infty$.}
\be
\nu_1^a q + \nu_2^a p = E \;.
\ee
If we normalise $q$ and $p$ by $E$, we recover that the area is precisely the area of the dual polygon to the toric fan of $B$. So the area of the surface of the set of points  
\be
\left\{ u\;,\;\left<u,\nu^a\right> \geq -1 \; \forall \;a \right\} \;.
\label{dualpol}
\ee 
For the example case of $B=\mathbb{P}^2$, this is illustrated in figure \ref{fig:p2tori}.

Comparing with (\ref{tripnE}), we therefore find that (in the limit $\mu \rightarrow \infty$) the triple intersection number is related to the area $A_B$ of the dual toric polygon of $B$ as
\be
\frac{r^3 \kappa_B}{2} = A_B  \;.
\label{rtoAfermimod}
\ee
For the example of $B=\mathbb{P}^2$ we have $A_B=\frac92$ and $r=3$ yielding $\kappa_B=\frac{1}{3}$.

\subsection{The cubic term for compact elliptic fibrations}
\label{sec:cubco}

In this section we propose that also for compact models there is a free Fermi gas picture from which the leading cubic term in the prepotential is emergent. The physics we have in mind is that for compact models there is still a constituent picture for the fields, and spacetime, in terms of some ultraviolet degrees of freedom. However, they do not behave as a free Fermi gas in general due to strong inter-particle interactions. Nonetheless, we propose that, at least in some cases, the high energy states of this Fermionic system do behave like a free Fermi gas. 

In this picture, we should be able to recover the cubic term in the prepotential as long as it is given by high energy modes. The cubic piece comes from the integral (\ref{JintE}). More specifically, it comes from the integration range
\be
\left.J\left(\mu\right)\right|_{\mathrm{cubic}} \subset \int_{\epsilon_{\mathrm{min}}}^{\mu} \frac{n(E)}{e^{E-\mu}+1} dE \;.
\label{JintEcub}
\ee
The integration range for energies $E > \mu$ always lead to terms with a factor exponential in the chemical potential. The simplest way to see this is to consider the case where $\mu < \epsilon_{\mathrm{min}}$ so that this integration regime is absent. In that case, we have the Fugacity expansion (\ref{Jorbifold}), which only has terms exponential in $\mu$. Physically, this is the orbifold regime, for which the cubic term is absent. 

From (\ref{JintEcub}), noting that $\mu$ caps the energy contribution, we see that the cubic term arises from high-energy modes only in the large chemical potential limit $\mu \rightarrow \infty$. This is precisely what we call infinite distances in field space. So approaching such infinite distances, we can recover the cubic term as emergent from the high-energy behaviour of the Fermi system, which we propose can sometimes be described as a free Fermi gas. 

The cases of compact models we would like to consider are elliptic fibrations over (reflexive) toric bases. So we just replace the cone over $B$ with a compact elliptic fibration. We refer to \cite{Chiang:1999tz}, for example, for details of such constructions. In these constructions, $T_B$ is now the Kahler modulus associated to the fibre volume, rather than a curve in the base. So it is Poincare dual to the base. Our proposal is that in the large distance limit $T_B \rightarrow \infty$, there is a free Fermi gas description (of the high-energy modes only) which yields the cubic term in $T_B$. 

The Fermi gas model we propose is simply the same model as (\ref{speccur}), with the vectors ${\bf \nu}^a$ being those of the toric fan of the base $B$ of the fibration. The relation between $T_B$ and $\mu$ is the same as in (\ref{j0f0cubic}) or (\ref{ttomuglobal}), with $r=1$.

To see that this model yields the correct triple intersection number, we first use the geometric relation (see, for example \cite{Chiang:1999tz})
\be
\left(J_B\right)^3 = \left(c_1(B)\right)^2 \;.
\label{jb3c12}
\ee
Here $J_B^3$ is the triple intersection of the Kahler form associated to $T_B$, and $c_1(B)$ is the first Chern class of the base.\footnote{The relation (\ref{jb3c12}) should be understood as integrating the six-form on the left over the whole Calabi-Yau, and the four-form on the right over the bases $B$.} We therefore have for the triple intersection number
\be
\kappa_B = \left(c_1(B)\right)^2 \;.
\ee

Next we note that $c_1(B)^2$ indeed computes (twice) the area of the dual polygon of the toric base $B$. 	
Given a lattice M, a lattice polytope $\Delta$ is a polytope in $\mathrm{M}\otimes \mathbb{R}$ with vertices in M. A reflexive polytope is a convex lattice polytope $\Delta$ with only the origin in its interior, such that its dual $\Delta^*$ defined through (\ref{dualpol}) is also a lattice polytope. For two-dimensional reflexive polygons, we have the following identity (see, for example \cite{Chiang:1999tz,He:2017gam}):
\be
c_1(B)^2 = 12 - \#\left(\partial\Delta\cap \mathrm{M}\right) \;,
\ee
with $\#\left(\partial\Delta\cap \mathrm{M}\right)$ the number of lattice points on the perimeter of $\Delta$, the polygon encoded by the vectors that we introduced in (\ref{vect}) for the base $B$. Let N be the dual lattice of M, then all reflexive polygons satisfy the following relation (twelve-point theorem \cite{2018arXiv180608351D}):
\be
\#\left(\partial\Delta\cap \mathrm{M}\right)+\#\left(\partial\Delta^*\cap \mathrm{N}\right) = 12 \;.
\ee
But by triangulation of $\Delta^*$ we can associate each lattice point on its boundary with a 2-simplex of area 1/2 inside it.
We therefore recover 
\be
\kappa_B = 2 A_B \;,
\ee
as predicted by the Fermi model (\ref{rtoAfermimod}). This holds for any reflexive toric base $B$.

We have therefore presented evidence that at least the cubic term in the prepotential can be understood as emergent from a free Fermi gas description also in compact models. We do not expect the exponential terms in the prepotential to be captured by this model, because they are not determined purely by high energy modes. In other words, the Fermi gas description is only valid for the high energy physics of the Fermi system.

\subsection{The cubic term for general one-parameter models}
\label{sec:spaintrel}

In section \ref{sec:intfermigas} we argued that the effective action is arising from integrating out the Fermi gas. There is also a four-dimensional spacetime integrating out interpretation of the exponential terms $e^{-T}$ in the genus-zero prepotential due to Gopakumar and Vafa \cite{Gopakumar:1998ii,Gopakumar:1998jq}. These are said to arise from integrating out $D2$-$D0$ bound states through Schwinger proper time integrals. Recall, as discussed in section \ref{sec:backgrsusy}, we can also think of integrating out the Fermi gas as a background field calculation, where we absorbed the graviphoton field strength $W$ into the string coupling. The difference between the approaches can be thought of as taking $g_s \rightarrow \infty$ in both cases, but in the Fermi gas one keeps $W$ finite, while in the GV approach one also sends $W \rightarrow 0$. 

The polynomial piece of the prepotential did not have an integrating out interpretation in the GV approach. However, in \cite{Hattab:2023moj} it was proposed that it can be understood as coming from integrating out ultraviolet degrees of freedom. The idea is to interpret the Mellin-Barnes representation of the period as a Schwinger integral, and then the analytic continuation of the Schwinger proper time allows an evaluation of the pole at the origin (which is the ultraviolet). This pole contribution then reproduces the polynomial piece, including the cubic term.

In this section we connect more explicitly the ideas in \cite{Hattab:2023moj} to the work in this paper. This will also provide a natural proposal for a generalisation of the Fermi Gas picture to any compact Calabi-Yau. 

The existence of the map (\ref{j0f0rel}) implies that we can expect a Mellin-Barnes representation for the classical piece of the Grand Canonical potential. For example, for the case $B = \mathbb{P}^1\times \mathbb{P}^1$ this is \cite{Marino:2011eh,Marino:2016new}:
\be
J_0(\mu) = -\frac12\int_{C}\frac{\text{d}z}{2\pi i} \frac{\Gamma(-z)\Gamma(z/2)^4}{\Gamma(z)}e^{\mu z} \;.
\label{MBrepj0}
\ee
The integration contour depends on the regime of the gas, as set by the chemical potential. The orbifold regime is when the chemical potential is less than the energy of the minimal energy eigenstate $\mu < \epsilon_{\mathrm{min}}=2 \log 2$, and then the integration contour is over the strictly positive poles. In the geometric regime, of large positive $\mu$, we instead take a contour around the zero pole and all the negative poles. The zero pole gives the polynomial piece of the Grand Canonical potential, and the negative poles give the exponentials. These are then mapped precisely to the period through the map (\ref{j0f0rel}).

In \cite{Hattab:2023moj} it was proposed to think of the Mellin-Barnes representations for the periods as capturing an integrating out calculation. With the complex parameter $z$ associated to an analytic continuation of the Schwinger proper time. It was then proposed that the pole at the origin, which yields the polynomial piece in the prepotential, is capturing integrating out the ultraviolet physics. It was further noted in \cite{Hattab:2023moj} that the integral representation is for the period(s), and not for the prepotential itself. This obstructed a direct map to the GV Schwinger integrals for $D2$-$D0$ branes, and instead the connection was indirect: the integrals contributed the exponentials to the prepotential, which is then matched onto the exponential terms in the periods after taking derivatives. 

In the context of the Fermi gas picture we have obtained a more detailed understanding of these relations. We have argued in section \ref{sec:intfermigas} that integrating out the Fermi Gas leads to the Grand Canonical potential. Therefore, indeed we can interpret the Mellin-Barnes representation (\ref{MBrepj0}) as capturing an integrating out calculation. However, the map to the effective action depends on the relation between $\mu$ and the moduli. As discussed in section \ref{sec:topadic}, this map relates the Grand Canonical potential to the Free energy (\ref{FtoJmap}) through (\ref{Tmumap}). More generally, we have the relation (\ref{ttomuglobal}). We see that it maps the contribution of the zero pole in the integral, the cubic piece in $\mu$, to the cubic piece in the genus-zero (supergravity) prepotential (as in (\ref{j0f0cubic})). Therefore, the zero pole is indeed capturing a direct integrating out calculation of the cubic piece. On the other hand, the exponential terms are mapped to non-perturbative terms in the string coupling, and cannot have a direct interpretation as integrating out $D2$-$D0$ branes.\footnote{For the specific case of the blow-up conifold, there is a GV type interpretation for these terms but in the context of the refined topological string: they arise in the Nekrasov-Shatashvili limit \cite{Nekrasov:2009rc}.} This is why they are contributing to the period rather than the prepotential. 

We can therefore see a general picture for the interpretation of the Mellin-Barnes integral representation of the periods, which can be applied naturally for any compact Calabi-Yau. We can think of this as an integrating out calculation for $J_0\left(\mu\right)$. Then, under a map which relates the Grand Canonical potential to the action, of type (\ref{Tmumap}), or more generally (\ref{ttomuglobal}), the zero pole indeed precisely gives the cubic piece in the prepotential. In this sense it arises from integrating out and is emergent. The exponential terms are also capturing terms in the effective action from integrating out, but they are non-perturbative in $g_s$, rather than worldsheet instantons which are classical from the string theory perspective. So from the spacetime four-dimensional perspective, emergence of the exponential terms in the genus-zero prepotential is simple to see via the GV approach of $D2$-$D0$ branes, and the emergence of the cubic piece is somewhat obstructed and subtle. From the Fermi gas perspective it is the opposite, emergence of the cubic piece is manifest, while the exponential terms in the genus-zero prepotential are difficult to see (they are non-perturbative in $\hbar$).

Understanding the Mellin-Barnes representation of the period as giving the Grand Canonical potential also shows that the chemical potential $\mu$ must be mapped to a global coordinate on the moduli space, since it exists in all regimes. Global pictures of the moduli space are best understood in the mirror type IIB setting, where the complex-structure moduli space is fully understood. In this setting, we should really identify $\mu$ with the parameters which appear in the definition of the Calabi-Yau as a hypersurface. 

We can consider, for example, the one-parameter model studied in \cite{Joshi:2019nzi,Palti:2021ubp,Bastian:2023shf,Hattab:2023moj}. The Calabi-Yau is given by the equations 
\bea
& & \frac{x_1^3}{3} + \frac{x_2^3}{3} + \frac{x_3^3}{3} + \psi x_1 x_2 x_3 = 0 \;, \nn \\
& & \frac{x_4^3}{3} + \frac{x_5^3}{3} + \frac{x_6^3}{3} + \psi x_4 x_5 x_6 = 0 \;,
\eea
where the $x_i$ are homogeneous coordinates in $\mathbb{P}^5$. The complex structure moduli space is spanned by the single (global) complex parameter $\psi$. The identification we propose with the chemical potential is then
\be
\mu = 2 \log \left(3\psi\right) \;,
\label{mupsimap}
\ee
which spans all the different regions in moduli space. 
The Mellin-Barnes integral representation for the period $\Pi_0\left(\psi\right)$ is then identified with the classical Grand Canonical potential, yielding a representation 
\begin{eqnarray}
\nonumber J_0(\mu)= -\int_C\frac{dz}{2\pi i}\frac{9e^{3\mu z}}{3^{6z}}&\bigg[&\frac{ \Gamma\left(\frac13 - z\right)^2 \Gamma(z)^4}{\Gamma\left(\frac13 + z\right)^2} +\frac{ \Gamma\left(\frac23 - z\right)^2 \Gamma(z)^4}{ \Gamma\left(\frac23 + z\right)^2} \\
& &- \frac{ (-1)^z (3i+\sqrt{3})\pi \Gamma\left(\frac23 - z\right) \Gamma(z)^4}{\Gamma\left(\frac13 + z\right) \Gamma(\frac23 + z)^2}
\nonumber \\
& & +\frac{(-1)^z (3i - \sqrt{3})\pi \Gamma\left(\frac13 - z\right) \Gamma(z)^4}{\Gamma\left(\frac13 + z\right)^2 \Gamma\left(\frac23 + z\right)}\bigg] \;.
\label{granp5}
\end{eqnarray}
The contour, as usual, needs to be chosen according to the value of $\mu$. For $\mu > 2\log 3$ we should include the negative integer poles and the origin, while for $\mu < 2\log 3$ we should include the strictly positive poles. The value $\epsilon_{\mathrm{min}}=2 \log 3$ should be interpreted as the energy of the minimal energy eigenstate of the system $\epsilon_{\mathrm{min}}$.

Note that the constant $r$ in (\ref{j0f0cubic}) is fixed by (\ref{mupsimap}) to $r=3$, since we have at large complex-structure/volume
\be
T_B = 6 \log \left(3\psi \right) = 3 \mu\;.
\label{classp5}
\ee
The exact coefficient in (\ref{mupsimap}) is fixed by demanding periodicity of $J(\mu)$ under $\mu \rightarrow \mu + 2\pi i$. In particular, the poles at $z=\frac13 + p$, with $p$ any positive integer, only lead to periodic terms for $r=3$.\footnote{The factor of $3$ can also been seen as the $\mathbb{Z}_3$ orbifold on the moduli space.}

While the classical map (\ref{mupsimap}) is exact, as discussed in section \ref{sec:glob}, it implies only a leading-order full quantum map, which in this case is therefore
\be
T_B = \frac{6\pi \mu}{\hbar} + ... \;.
\label{quanp5}
\ee

To summarise, we have proposed a Grand Canonical potential (\ref{granp5}), and a leading-order quantum map (\ref{quanp5}), for this example compact Calabi-Yau. But the procedure is completely general: the Grand Canonical potential is identified with the period, and the classical map between the chemical potential and the complex-structure parameter is determined by demanding periodicity of the Grand Canonical potential under $\mu \rightarrow \mu + 2\pi i$. Then the leading-order quantum map is determined from the classical map.

The interpretation of this is that we should understand the cubic term as arising from integrating out some ultraviolet degrees of freedom with an associated Grand Canonical potential $J(\mu)$. Those degrees of freedom need not be described by a free Fermi gas, indeed we did not specify any Hamiltonian but only a Grand Canonical potential. In this sense, our analysis in this section is more general than the elliptic fibrations one in section \ref{sec:cubco}, but is less explicit in identifying the ultraviolet degrees of freedom. It would be interesting to understand if it is possible to determine a Hamiltonian for the ultraviolet degrees of freedom starting from the period Mellin-Barnes representation. 

\section{Summary and discussion on Swampland towers}
\label{sec:disc}

In this paper we studied the Emergence Proposal in the context of type II string theory compactified on Calabi-Yau manifolds. The key question is regarding whether the prepotential arises from integrating out ultraviolet states. It has long been known that the exponential terms, the worldsheet instanton contributions, can be understood as arising from integrating out $D2$-$D0$ branes \cite{Gopakumar:1998ii,Gopakumar:1998jq}. More recently, the work in \cite{Hattab:2023moj} proposed that the tree-level cubic part of the prepotential arises from integrating out point-like ultraviolet degrees of freedom. The $D2$-$D0$ states would then be understood as made from these constituent degrees of freedom, of which there are only a few. In this paper we presented further evidence for this picture, with concrete models of these ultraviolet degrees of freedom. 

The work was motivated by the results of \cite{Marino:2011eh,Grassi:2014zfa}, which showed that topological strings on certain non-compact Calabi-Yau cones can be understood in terms of Fermi gases. While in those works the map to Fermi gases was somewhat formal, forming a model of eigenvalues of a matrix model, we proposed that this is actually capturing emergence: so the Fermion in the gas is (at least part of) the ultraviolet degree(s) of freedom discussed above. We presented evidence for this by considering a toy model for a simple free two-dimensional Fermion. The model led to the interpretation that integrating out the Fermi gas induces dynamics for the chemical potential $\mu$ of the system.

In this integrating out picture, we identify the Grand Canonical potential of the ultraviolet system $J(\mu)$ with the effective action for $\mu$. In the context of topological strings, this effective action can be mapped to the Free Energy. The Fermi gas model, for some settings, indeed reproduces the topological string Free energy to all orders (including non-perturbative terms) in the string coupling. To see this identification, one needs a map between the chemical potential $\mu$ and the string moduli $T$. This map is a `quantum map' in that it involves the Planck constant of the Fermi gas.

To make contact with the Swampland and dynamical gravity, we generalised the Fermi gas picture to compact Calabi-Yau manifolds. In these cases, the understanding of the ultraviolet physics was less complete, but the elements required for extracting the cubic term in the prepotential were presented. For the case of compact elliptic fibrations over reflexive toric bases this information amounted to the high energy behaviour of the ultraviolet degrees of freedom, which we proposed is that of a Fermi gas. While for the general one-parameter case the information could be deduced by studying a so-called `classical' map between the Grand Canonical potential and the period. 

It would be very interesting to understand how to generalise the results to more compact cases. We expect that this would involve interacting, rather than free, theories which modify also the high-energy behaviour. For example, in \cite{Marino:2012az} it was shown that some interactions do change the coefficient in front of the cubic term. 

Local mirror symmetry plays an important role in determining the Hamiltonian of the Fermi gas. This is the same role that it plays in determining the spectral curve of topological strings. There is a very general formulation of local mirror symmetry in terms of degenerations of mixed Hodge structures. Indeed, this general formulation played a role in classifying and understanding infinite distances in field space in \cite{Grimm:2018ohb,Grimm:2018cpv}. It would be interesting to see if there is similarly a general formulation of the Fermi gas model in terms of Hodge structure data.

The Fermi gas description is closely related to the description of the topological string partition function in terms of a melting crystal \cite{Okounkov:2003sp,Iqbal:2003ds,Ooguri:2009ijd,Ooguri:2009ri}. We expect that this should also provide insights into the Emergence and integrating out picture. It would be interesting to revisit these ideas in the context of the crystal melting framework.

\subsection*{The towers of states of the Swampland conjectures}

The Emergence Proposal can be the general physics underlying many Swampland conjectures \cite{Palti:2019pca}. We can therefore consider the implications of the results of this paper on Swampland conjectures. In particular, the relevant conjectures here are the Distance Conjecture \cite{Ooguri:2006in} (and its refined version \cite{Klaewer:2016kiy}), and the magnetic Weak Gravity Conjecture \cite{Arkani-Hamed:2006emk}. 

We note that from a mathematical rigour perspective, the Distance Conjecture is already firmly established in the type II ${\cal N}=2$ setting \cite{Palti:2017elp,Grimm:2018ohb,Grimm:2018cpv,Corvilain:2018lgw}. In particular, in \cite{Grimm:2018ohb} a mathematical proof was presented for the full complex-structure moduli space of Calabi-Yau manifolds. The proof holds up an assumption of the population of the BPS states that become light.\footnote{In \cite{Palti:2021ubp}, it was shown that if the charged BPS black holes exist then they would populate these BPS states. It is also possible to see that the BPS states would be present at strong string coupling where they would be associated with the Kaluza-Klein modes of the M-theory circle. Since the BPS spectrum (index) should not change in going from strong to weak coupling, the states should also be populated at weak coupling. We also know that the KK modes are electrically charged under the graviphoton, and so are light relative to the Planck scale, and therefore a tower of the electric/light BPS states at weak coupling must be populated. This gives a simple argument for the population of the tower of BPS states.} 
 
We are therefore not able to offer an advance in rigour, but rather conceptual insights into the Distance Conjecture. This can be useful to understand its general underlying microscopic physics, even away from the type II Calabi-Yau setting. Making contact with the towers of the Distance conjecture explicitly is not simple. This is because they are formulated in the language of weak string coupling, while the Fermi gas picture is simple only at strong string coupling. In other words, we would like to understand the spectrum of $D2$-$D0$ states, but these are composite non-perturbative objects from the Fermi gas perspective. 

A simple starting point to probe the $D2$-$D0$ spectrum is by its contribution to the genus-zero prepotential after integrating out. The $D2$-$D0$ bound states give rise to the exponential terms, while the pure $D0$ states give rise to the $\zeta(3)\chi$ term, with $\chi$ being the Euler number of the Calabi-Yau.\footnote{Note that these contributions can be absent in some cases, \cite{Ferrara:1995yx,Palti:2020qlc} (due to some higher supersymmetry). Therefore such a probe cannot be a general one for the ultraviolet physics.} In the Fermi gas model we can identify the exponentials with non-perturbative  contributions to the Grand Canonical potential behaving as $e^{-\frac{\mu}{\hbar}}$. 

The $D0$ contribution can be seen from the part of the Grand Canonical potential that is independent of $\mu$, as given by (\ref{indepmu}) for the blown-up conifold case. We can see in (\ref{indepmu}) the contribution (going as $\zeta(3)$) at strong string coupling, small $\hbar$, to the genus-zero prepotential within the classical piece $J_0(\mu)$ which goes as $\frac{1}{\hbar}$. This contribution is from the classical map (\ref{j0f0rel}). We can also see the weak string coupling, large $\hbar$, contribution relevant for the full quantum map (\ref{Tmumap}), which behaves as $\hbar^2$. This shows that we have a precise identification of the $D0$ tower, which is the leading lightest tower, also in the ultraviolet Fermi gas picture. 

Within emergence, the towers of states in the infrared are required to drive the kinetic terms to strong coupling in the ultraviolet, where a dual description becomes necessary. For our cases, the Fermi gas model is (a part of) this dual description. In this sense, identifying a strongly-coupled description as relevant for the ultraviolet predicts the towers of states. In order to understand better the nature of the towers from the ultraviolet perspective we would like to have a microscopic picture of these composite states which is more direct than their influence on the Grand Canonical potential from integrating out. We aim to investigate such a microscopic picture in future work.

\vspace{0.1cm}
{\bf Acknowledgements}
\noindent
We would like to thank Ralph Blumenhagen, Thomas Grimm, Marcos Marino and Timo Weigand for very useful discussions. The work of JH and EP is supported by the Israel Science Foundation (grant No. 741/20) and by the German Research Foundation through a German-Israeli Project Cooperation (DIP) grant ``Holography and the Swampland". The work of EP is supported by the Israel planning and budgeting committee grant for supporting theoretical high energy physics.

\appendix

\bibliographystyle{jhep}
\bibliography{Higuchi}

\providecommand{\href}[2]{#2}\begingroup\raggedright\begin{thebibliography}{10}

\bibitem{tHooft:1993dmi}
G.~'t~Hooft, \emph{{Dimensional reduction in quantum gravity}}, {\emph{Conf. Proc. C} {\bfseries 930308} (1993) 284} [\href{https://arxiv.org/abs/gr-qc/9310026}{{\ttfamily gr-qc/9310026}}].

\bibitem{Palti:2019pca}
E.~Palti, \emph{{The Swampland: Introduction and Review}}, \href{https://doi.org/10.1002/prop.201900037}{\emph{Fortsch. Phys.} {\bfseries 67} (2019) 1900037} [\href{https://arxiv.org/abs/1903.06239}{{\ttfamily 1903.06239}}].

\bibitem{Harlow:2015lma}
D.~Harlow, \emph{{Wormholes, Emergent Gauge Fields, and the Weak Gravity Conjecture}}, \href{https://doi.org/10.1007/JHEP01(2016)122}{\emph{JHEP} {\bfseries 01} (2016) 122} [\href{https://arxiv.org/abs/1510.07911}{{\ttfamily 1510.07911}}].

\bibitem{Heidenreich:2017sim}
B.~Heidenreich, M.~Reece and T.~Rudelius, \emph{{The Weak Gravity Conjecture and Emergence from an Ultraviolet Cutoff}}, \href{https://doi.org/10.1140/epjc/s10052-018-5811-3}{\emph{Eur. Phys. J. C} {\bfseries 78} (2018) 337} [\href{https://arxiv.org/abs/1712.01868}{{\ttfamily 1712.01868}}].

\bibitem{Grimm:2018ohb}
T.W.~Grimm, E.~Palti and I.~Valenzuela, \emph{{Infinite Distances in Field Space and Massless Towers of States}}, \href{https://doi.org/10.1007/JHEP08(2018)143}{\emph{JHEP} {\bfseries 08} (2018) 143} [\href{https://arxiv.org/abs/1802.08264}{{\ttfamily 1802.08264}}].

\bibitem{Heidenreich:2018kpg}
B.~Heidenreich, M.~Reece and T.~Rudelius, \emph{{Emergence of Weak Coupling at Large Distance in Quantum Gravity}}, \href{https://doi.org/10.1103/PhysRevLett.121.051601}{\emph{Phys. Rev. Lett.} {\bfseries 121} (2018) 051601} [\href{https://arxiv.org/abs/1802.08698}{{\ttfamily 1802.08698}}].

\bibitem{Grimm:2018cpv}
T.W.~Grimm, C.~Li and E.~Palti, \emph{{Infinite Distance Networks in Field Space and Charge Orbits}}, \href{https://doi.org/10.1007/JHEP03(2019)016}{\emph{JHEP} {\bfseries 03} (2019) 016} [\href{https://arxiv.org/abs/1811.02571}{{\ttfamily 1811.02571}}].

\bibitem{Corvilain:2018lgw}
P.~Corvilain, T.W.~Grimm and I.~Valenzuela, \emph{{The Swampland Distance Conjecture for K\"ahler moduli}}, \href{https://doi.org/10.1007/JHEP08(2019)075}{\emph{JHEP} {\bfseries 08} (2019) 075} [\href{https://arxiv.org/abs/1812.07548}{{\ttfamily 1812.07548}}].

\bibitem{Lee:2019xtm}
S.-J.~Lee, W.~Lerche and T.~Weigand, \emph{{Emergent strings, duality and weak coupling limits for two-form fields}}, \href{https://doi.org/10.1007/JHEP02(2022)096}{\emph{JHEP} {\bfseries 02} (2022) 096} [\href{https://arxiv.org/abs/1904.06344}{{\ttfamily 1904.06344}}].

\bibitem{Lee:2019wij}
S.-J.~Lee, W.~Lerche and T.~Weigand, \emph{{Emergent strings from infinite distance limits}}, \href{https://doi.org/10.1007/JHEP02(2022)190}{\emph{JHEP} {\bfseries 02} (2022) 190} [\href{https://arxiv.org/abs/1910.01135}{{\ttfamily 1910.01135}}].

\bibitem{Palti:2020tsy}
E.~Palti, \emph{{Fermions and the Swampland}}, \href{https://doi.org/10.1016/j.physletb.2020.135617}{\emph{Phys. Lett. B} {\bfseries 808} (2020) 135617} [\href{https://arxiv.org/abs/2005.08538}{{\ttfamily 2005.08538}}].

\bibitem{Marchesano:2022avb}
F.~Marchesano and M.~Wiesner, \emph{{4d strings at strong coupling}}, \href{https://doi.org/10.1007/JHEP08(2022)004}{\emph{JHEP} {\bfseries 08} (2022) 004} [\href{https://arxiv.org/abs/2202.10466}{{\ttfamily 2202.10466}}].

\bibitem{Castellano:2022bvr}
A.~Castellano, A.~Herr\'aez and L.E.~Ib\'a\~nez, \emph{{The emergence proposal in quantum gravity and the species scale}}, \href{https://doi.org/10.1007/JHEP06(2023)047}{\emph{JHEP} {\bfseries 06} (2023) 047} [\href{https://arxiv.org/abs/2212.03908}{{\ttfamily 2212.03908}}].

\bibitem{vandeHeisteeg:2022btw}
D.~van~de Heisteeg, C.~Vafa, M.~Wiesner and D.H.~Wu, \emph{{Moduli-dependent Species Scale}},  \href{https://arxiv.org/abs/2212.06841}{{\ttfamily 2212.06841}}.

\bibitem{Cota:2022maf}
C.F.~Cota, A.~Mininno, T.~Weigand and M.~Wiesner, \emph{{The asymptotic weak gravity conjecture in M-theory}}, \href{https://doi.org/10.1007/JHEP08(2023)057}{\emph{JHEP} {\bfseries 08} (2023) 057} [\href{https://arxiv.org/abs/2212.09758}{{\ttfamily 2212.09758}}].

\bibitem{Cribiori:2022nke}
N.~Cribiori, D.~L\"ust and G.~Staudt, \emph{{Black hole entropy and moduli-dependent species scale}}, \href{https://doi.org/10.1016/j.physletb.2023.138113}{\emph{Phys. Lett. B} {\bfseries 844} (2023) 138113} [\href{https://arxiv.org/abs/2212.10286}{{\ttfamily 2212.10286}}].

\bibitem{Marchesano:2022axe}
F.~Marchesano and L.~Melotti, \emph{{EFT strings and emergence}}, \href{https://doi.org/10.1007/JHEP02(2023)112}{\emph{JHEP} {\bfseries 02} (2023) 112} [\href{https://arxiv.org/abs/2211.01409}{{\ttfamily 2211.01409}}].

\bibitem{Blumenhagen:2023tev}
R.~Blumenhagen, N.~Cribiori, A.~Gligovic and A.~Paraskevopoulou, \emph{{Demystifying the Emergence Proposal}},  \href{https://arxiv.org/abs/2309.11551}{{\ttfamily 2309.11551}}.

\bibitem{Castellano:2023qhp}
A.~Castellano, A.~Herr\'aez and L.E.~Ib\'a\~nez, \emph{{Towers and hierarchies in the Standard Model from Emergence in Quantum Gravity}}, \href{https://doi.org/10.1007/JHEP10(2023)172}{\emph{JHEP} {\bfseries 10} (2023) 172} [\href{https://arxiv.org/abs/2302.00017}{{\ttfamily 2302.00017}}].

\bibitem{Burgess:2023pnk}
C.P.~Burgess and F.~Quevedo, \emph{{Perils of towers in the swamp: dark dimensions and the robustness of EFTs}}, \href{https://doi.org/10.1007/JHEP09(2023)159}{\emph{JHEP} {\bfseries 09} (2023) 159} [\href{https://arxiv.org/abs/2304.03902}{{\ttfamily 2304.03902}}].

\bibitem{Baume:2023msm}
F.~Baume and J.~Calder\'on-Infante, \emph{{On Higher-Spin Points and Infinite Distances in Conformal Manifolds}},  \href{https://arxiv.org/abs/2305.05693}{{\ttfamily 2305.05693}}.

\bibitem{Cribiori:2023ffn}
N.~Cribiori, D.~Lust and C.~Montella, \emph{{Species entropy and thermodynamics}}, \href{https://doi.org/10.1007/JHEP10(2023)059}{\emph{JHEP} {\bfseries 10} (2023) 059} [\href{https://arxiv.org/abs/2305.10489}{{\ttfamily 2305.10489}}].

\bibitem{Blumenhagen:2023yws}
R.~Blumenhagen, A.~Gligovic and A.~Paraskevopoulou, \emph{{The emergence proposal and the emergent string}}, \href{https://doi.org/10.1007/JHEP10(2023)145}{\emph{JHEP} {\bfseries 10} (2023) 145} [\href{https://arxiv.org/abs/2305.10490}{{\ttfamily 2305.10490}}].

\bibitem{Blumenhagen:2023xmk}
R.~Blumenhagen, N.~Cribiori, A.~Gligovic and A.~Paraskevopoulou, \emph{{The Emergent M-theory Limit}},  \href{https://arxiv.org/abs/2309.11554}{{\ttfamily 2309.11554}}.

\bibitem{Seo:2023xsb}
M.-S.~Seo, \emph{{(In)stability of de Sitter vacuum in light of distance conjecture and emergence proposal}}, \href{https://doi.org/10.1007/JHEP09(2023)031}{\emph{JHEP} {\bfseries 09} (2023) 031} [\href{https://arxiv.org/abs/2305.18673}{{\ttfamily 2305.18673}}].

\bibitem{Calderon-Infante:2023ler}
J.~Calder\'on-Infante, A.~Castellano, A.~Herr\'aez and L.E.~Ib\'a\~nez, \emph{{Entropy Bounds and the Species Scale Distance Conjecture}},  \href{https://arxiv.org/abs/2306.16450}{{\ttfamily 2306.16450}}.

\bibitem{DeBiasio:2023hzo}
D.~De~Biasio, \emph{{Geometric flows and the Swampland}},  \href{https://arxiv.org/abs/2307.08320}{{\ttfamily 2307.08320}}.

\bibitem{Calderon-Infante:2023uhz}
J.~Calder\'on-Infante, M.~Delgado and A.M.~Uranga, \emph{{Emergence of Species Scale Black Hole Horizons}},  \href{https://arxiv.org/abs/2310.04488}{{\ttfamily 2310.04488}}.

\bibitem{Castellano:2023aum}
A.~Castellano, A.~Herr\'aez and L.E.~Ib\'a\~nez, \emph{{On the Species Scale, Modular Invariance and the Gravitational EFT expansion}},  \href{https://arxiv.org/abs/2310.07708}{{\ttfamily 2310.07708}}.

\bibitem{Castellano:2023jjt}
A.~Castellano, I.~Ruiz and I.~Valenzuela, \emph{{Stringy Evidence for a Universal Pattern at Infinite Distance}},  \href{https://arxiv.org/abs/2311.01536}{{\ttfamily 2311.01536}}.

\bibitem{Castellano:2023stg}
A.~Castellano, I.~Ruiz and I.~Valenzuela, \emph{{A Universal Pattern in Quantum Gravity at Infinite Distance}},  \href{https://arxiv.org/abs/2311.01501}{{\ttfamily 2311.01501}}.

\bibitem{Marchesano:2023thx}
F.~Marchesano, L.~Melotti and L.~Paoloni, \emph{{On the moduli space curvature at infinity}},  \href{https://arxiv.org/abs/2311.07979}{{\ttfamily 2311.07979}}.

\bibitem{Basile:2023blg}
I.~Basile, D.~Lust and C.~Montella, \emph{{Shedding black hole light on the emergent string conjecture}},  \href{https://arxiv.org/abs/2311.12113}{{\ttfamily 2311.12113}}.

\bibitem{Cota:2023uir}
C.F.~Cota, A.~Mininno, T.~Weigand and M.~Wiesner, \emph{{The Minimal Weak Gravity Conjecture}},  \href{https://arxiv.org/abs/2312.04619}{{\ttfamily 2312.04619}}.

\bibitem{Hattab:2023moj}
J.~Hattab and E.~Palti, \emph{{On the particle picture of Emergence}},  \href{https://arxiv.org/abs/2312.15440}{{\ttfamily 2312.15440}}.

\bibitem{Casas:2024ttx}
G.F.~Casas, L.E.~Ib\'a\~nez and F.~Marchesano, \emph{{Yukawa Couplings at Infinite Distance and Swampland Towers in Chiral Theories}},  \href{https://arxiv.org/abs/2403.09775}{{\ttfamily 2403.09775}}.

\bibitem{Blumenhagen:2024ydy}
R.~Blumenhagen, N.~Cribiori, A.~Gligovic and A.~Paraskevopoulou, \emph{{Emergence of $R^4$-terms in M-theory}},  \href{https://arxiv.org/abs/2404.01371}{{\ttfamily 2404.01371}}.

\bibitem{Gopakumar:1998ii}
R.~Gopakumar and C.~Vafa, \emph{{M theory and topological strings. 1.}},  \href{https://arxiv.org/abs/hep-th/9809187}{{\ttfamily hep-th/9809187}}.

\bibitem{Gopakumar:1998jq}
R.~Gopakumar and C.~Vafa, \emph{{M theory and topological strings. 2.}},  \href{https://arxiv.org/abs/hep-th/9812127}{{\ttfamily hep-th/9812127}}.

\bibitem{Marino:2011eh}
M.~Marino and P.~Putrov, \emph{{ABJM theory as a Fermi gas}}, \href{https://doi.org/10.1088/1742-5468/2012/03/P03001}{\emph{J. Stat. Mech.} {\bfseries 1203} (2012) P03001} [\href{https://arxiv.org/abs/1110.4066}{{\ttfamily 1110.4066}}].

\bibitem{Aharony:2008ug}
O.~Aharony, O.~Bergman, D.L.~Jafferis and J.~Maldacena, \emph{{N=6 superconformal Chern-Simons-matter theories, M2-branes and their gravity duals}}, \href{https://doi.org/10.1088/1126-6708/2008/10/091}{\emph{JHEP} {\bfseries 10} (2008) 091} [\href{https://arxiv.org/abs/0806.1218}{{\ttfamily 0806.1218}}].

\bibitem{Aganagic:2002wv}
M.~Aganagic, A.~Klemm, M.~Marino and C.~Vafa, \emph{{Matrix model as a mirror of Chern-Simons theory}}, \href{https://doi.org/10.1088/1126-6708/2004/02/010}{\emph{JHEP} {\bfseries 02} (2004) 010} [\href{https://arxiv.org/abs/hep-th/0211098}{{\ttfamily hep-th/0211098}}].

\bibitem{Marino:2009jd}
M.~Marino and P.~Putrov, \emph{{Exact Results in ABJM Theory from Topological Strings}}, \href{https://doi.org/10.1007/JHEP06(2010)011}{\emph{JHEP} {\bfseries 06} (2010) 011} [\href{https://arxiv.org/abs/0912.3074}{{\ttfamily 0912.3074}}].

\bibitem{Drukker:2010nc}
N.~Drukker, M.~Marino and P.~Putrov, \emph{{From weak to strong coupling in ABJM theory}}, \href{https://doi.org/10.1007/s00220-011-1253-6}{\emph{Commun. Math. Phys.} {\bfseries 306} (2011) 511} [\href{https://arxiv.org/abs/1007.3837}{{\ttfamily 1007.3837}}].

\bibitem{Gopakumar:1998ki}
R.~Gopakumar and C.~Vafa, \emph{{On the gauge theory / geometry correspondence}}, \href{https://doi.org/10.4310/ATMP.1999.v3.n5.a5}{\emph{Adv. Theor. Math. Phys.} {\bfseries 3} (1999) 1415} [\href{https://arxiv.org/abs/hep-th/9811131}{{\ttfamily hep-th/9811131}}].

\bibitem{Grassi:2014zfa}
A.~Grassi, Y.~Hatsuda and M.~Marino, \emph{{Topological Strings from Quantum Mechanics}}, \href{https://doi.org/10.1007/s00023-016-0479-4}{\emph{Annales Henri Poincare} {\bfseries 17} (2016) 3177} [\href{https://arxiv.org/abs/1410.3382}{{\ttfamily 1410.3382}}].

\bibitem{Hatsuda:2013oxa}
Y.~Hatsuda, M.~Marino, S.~Moriyama and K.~Okuyama, \emph{{Non-perturbative effects and the refined topological string}}, \href{https://doi.org/10.1007/JHEP09(2014)168}{\emph{JHEP} {\bfseries 09} (2014) 168} [\href{https://arxiv.org/abs/1306.1734}{{\ttfamily 1306.1734}}].

\bibitem{Marino:2016new}
M.~Marino, \emph{{Localization at large N in Chern\textendash{}Simons-matter theories}}, \href{https://doi.org/10.1088/1751-8121/aa5f69}{\emph{J. Phys. A} {\bfseries 50} (2017) 443007} [\href{https://arxiv.org/abs/1608.02959}{{\ttfamily 1608.02959}}].

\bibitem{Okounkov:2003sp}
A.~Okounkov, N.~Reshetikhin and C.~Vafa, \emph{{Quantum Calabi-Yau and classical crystals}}, \href{https://doi.org/10.1007/0-8176-4467-9_16}{\emph{Prog. Math.} {\bfseries 244} (2006) 597} [\href{https://arxiv.org/abs/hep-th/0309208}{{\ttfamily hep-th/0309208}}].

\bibitem{Iqbal:2003ds}
A.~Iqbal, N.~Nekrasov, A.~Okounkov and C.~Vafa, \emph{{Quantum foam and topological strings}}, \href{https://doi.org/10.1088/1126-6708/2008/04/011}{\emph{JHEP} {\bfseries 04} (2008) 011} [\href{https://arxiv.org/abs/hep-th/0312022}{{\ttfamily hep-th/0312022}}].

\bibitem{Ooguri:2009ijd}
H.~Ooguri and M.~Yamazaki, \emph{{Crystal Melting and Toric Calabi-Yau Manifolds}}, \href{https://doi.org/10.1007/s00220-009-0836-y}{\emph{Commun. Math. Phys.} {\bfseries 292} (2009) 179} [\href{https://arxiv.org/abs/0811.2801}{{\ttfamily 0811.2801}}].

\bibitem{Ooguri:2009ri}
H.~Ooguri and M.~Yamazaki, \emph{{Emergent Calabi-Yau Geometry}}, \href{https://doi.org/10.1103/PhysRevLett.102.161601}{\emph{Phys. Rev. Lett.} {\bfseries 102} (2009) 161601} [\href{https://arxiv.org/abs/0902.3996}{{\ttfamily 0902.3996}}].

\bibitem{Kapusta_Gale_2006}
J.I.~Kapusta and C.~Gale, \emph{Finite-Temperature Field Theory: Principles and Applications}, Cambridge Monographs on Mathematical Physics, Cambridge University Press, 2~ed. (2006).

\bibitem{Dedushenko:2014nya}
M.~Dedushenko and E.~Witten, \emph{{Some Details On The Gopakumar-Vafa and Ooguri-Vafa Formulas}}, \href{https://doi.org/10.4310/ATMP.2016.v20.n1.a1}{\emph{Adv. Theor. Math. Phys.} {\bfseries 20} (2016) 1} [\href{https://arxiv.org/abs/1411.7108}{{\ttfamily 1411.7108}}].

\bibitem{Chiang:1999tz}
T.M.~Chiang, A.~Klemm, S.-T.~Yau and E.~Zaslow, \emph{{Local mirror symmetry: Calculations and interpretations}}, \href{https://doi.org/10.4310/ATMP.1999.v3.n3.a3}{\emph{Adv. Theor. Math. Phys.} {\bfseries 3} (1999) 495} [\href{https://arxiv.org/abs/hep-th/9903053}{{\ttfamily hep-th/9903053}}].

\bibitem{He:2017gam}
Y.-H.~He, R.-K.~Seong and S.-T.~Yau, \emph{{Calabi\textendash{}Yau Volumes and Reflexive Polytopes}}, \href{https://doi.org/10.1007/s00220-018-3128-6}{\emph{Commun. Math. Phys.} {\bfseries 361} (2018) 155} [\href{https://arxiv.org/abs/1704.03462}{{\ttfamily 1704.03462}}].

\bibitem{2018arXiv180608351D}
D.I.~{Dais}, \emph{{On the Twelve-Point Theorem for $\ell$-Reflexive Polygons}}, \href{https://doi.org/10.48550/arXiv.1806.08351}{\emph{arXiv e-prints} (2018) arXiv:1806.08351} [\href{https://arxiv.org/abs/1806.08351}{{\ttfamily 1806.08351}}].

\bibitem{Nekrasov:2009rc}
N.A.~Nekrasov and S.L.~Shatashvili, \emph{{Quantization of Integrable Systems and Four Dimensional Gauge Theories}},  in \emph{{16th International Congress on Mathematical Physics}}, pp.~265--289, 2010, \href{https://doi.org/10.1142/9789814304634_0015}{DOI} [\href{https://arxiv.org/abs/0908.4052}{{\ttfamily 0908.4052}}].

\bibitem{Joshi:2019nzi}
A.~Joshi and A.~Klemm, \emph{{Swampland Distance Conjecture for One-Parameter Calabi-Yau Threefolds}}, \href{https://doi.org/10.1007/JHEP08(2019)086}{\emph{JHEP} {\bfseries 08} (2019) 086} [\href{https://arxiv.org/abs/1903.00596}{{\ttfamily 1903.00596}}].

\bibitem{Palti:2021ubp}
E.~Palti, \emph{{Stability of BPS states and weak coupling limits}}, \href{https://doi.org/10.1007/JHEP08(2021)091}{\emph{JHEP} {\bfseries 08} (2021) 091} [\href{https://arxiv.org/abs/2107.01539}{{\ttfamily 2107.01539}}].

\bibitem{Bastian:2023shf}
B.~Bastian, D.~van~de Heisteeg and L.~Schlechter, \emph{{Beyond Large Complex Structure: Quantized Periods and Boundary Data for One-Modulus Singularities}},  \href{https://arxiv.org/abs/2306.01059}{{\ttfamily 2306.01059}}.

\bibitem{Marino:2012az}
M.~Mari\~no and P.~Putrov, \emph{{Interacting fermions and N=2 Chern-Simons-matter theories}}, \href{https://doi.org/10.1007/JHEP11(2013)199}{\emph{JHEP} {\bfseries 11} (2013) 199} [\href{https://arxiv.org/abs/1206.6346}{{\ttfamily 1206.6346}}].

\bibitem{Ooguri:2006in}
H.~Ooguri and C.~Vafa, \emph{{On the Geometry of the String Landscape and the Swampland}}, \href{https://doi.org/10.1016/j.nuclphysb.2006.10.033}{\emph{Nucl. Phys. B} {\bfseries 766} (2007) 21} [\href{https://arxiv.org/abs/hep-th/0605264}{{\ttfamily hep-th/0605264}}].

\bibitem{Klaewer:2016kiy}
D.~Klaewer and E.~Palti, \emph{{Super-Planckian Spatial Field Variations and Quantum Gravity}}, \href{https://doi.org/10.1007/JHEP01(2017)088}{\emph{JHEP} {\bfseries 01} (2017) 088} [\href{https://arxiv.org/abs/1610.00010}{{\ttfamily 1610.00010}}].

\bibitem{Arkani-Hamed:2006emk}
N.~Arkani-Hamed, L.~Motl, A.~Nicolis and C.~Vafa, \emph{{The String landscape, black holes and gravity as the weakest force}}, \href{https://doi.org/10.1088/1126-6708/2007/06/060}{\emph{JHEP} {\bfseries 06} (2007) 060} [\href{https://arxiv.org/abs/hep-th/0601001}{{\ttfamily hep-th/0601001}}].

\bibitem{Palti:2017elp}
E.~Palti, \emph{{The Weak Gravity Conjecture and Scalar Fields}}, \href{https://doi.org/10.1007/JHEP08(2017)034}{\emph{JHEP} {\bfseries 08} (2017) 034} [\href{https://arxiv.org/abs/1705.04328}{{\ttfamily 1705.04328}}].

\bibitem{Ferrara:1995yx}
S.~Ferrara, J.A.~Harvey, A.~Strominger and C.~Vafa, \emph{{Second quantized mirror symmetry}}, \href{https://doi.org/10.1016/0370-2693(95)01074-Z}{\emph{Phys. Lett. B} {\bfseries 361} (1995) 59} [\href{https://arxiv.org/abs/hep-th/9505162}{{\ttfamily hep-th/9505162}}].

\bibitem{Palti:2020qlc}
E.~Palti, C.~Vafa and T.~Weigand, \emph{{Supersymmetric Protection and the Swampland}}, \href{https://doi.org/10.1007/JHEP06(2020)168}{\emph{JHEP} {\bfseries 06} (2020) 168} [\href{https://arxiv.org/abs/2003.10452}{{\ttfamily 2003.10452}}].

\end{thebibliography}\endgroup

\end{document}